\documentstyle[11pt,newpasp,twoside]{article}
\def\arg#1{{\it#1\/}}

\def\edcomment#1{\iffalse\marginpar{\raggedright\sl#1\/}\else\relax\fi}
\reversemarginpar

\begin{document}
\title{The Future of High Angular Resolution Star and \\
       Planet Formation Science in the Optical/Infrared}
 \author{Lynne A. Hillenbrand}
\affil{California Institute of Technology, Pasadena, CA 91125}

\begin{abstract}
 
This presentation summarizes how some of the most pressing questions 
in the field of star and planet formation can  be addressed by 
high angular resolution optical/infrared capabilities, and how many of
these capabilities will in fact be available with realization of
the space and ground facilities currently being planned for the 2005-2020 
time frame.
 
\end{abstract}

\section{Introduction}

The quest to understand the formation of stars and planets, especially exo-solar
planetary systems
showing some resemblance to our own Sun and solar system, is a fundamental goal of modern 
astrophysics. Star and planet formation studies make use of all observational techniques 
at all wavelengths ranging from x-ray to radio bands.  This presentation focuses on 
optical/infrared questions and capabilities involving high spatial resolution techniques.

There are a number of problems in star and planet formation that would benefit
substantially from improved spatial resolution over current capabilities
in the optical and infrared.  Here, I focus on three.

\underline{The initial mass function (IMF)}
The initial mass function, or spectrum of stellar/sub-stellar masses produced
in the process of molecular cloud fragmentation is one of the most fundamental products
of the star formation process.  Rigorous observational effort over the 
last decade has provided insight into the form and detailed shape of the IMF 
at the point of ``birth" of stellar clusters, all the way from the 
most massive stars through to sub-stellar mass objects. 
While the IMF rises with an approximately Salpeter slope from $>$100 M$_\odot$
to $<$2 M$_\odot$ there is now ample evidence for a change in slope between 0.5-2 M$_\odot$
and another more dramatic change below 0.5 $M_\odot$.  
A function in which the slope is continually changing with
mass, such as the log-normal form, probably best describes the data in hand.
Outstanding IMF issues include the maximum and minimum masses formed
by the star formation process; the variation (or lack thereof)
in dN/dM as a function of metallicity, temperature/pressure, or
local stellar density; and the binary frequency and binary mass ratio
(q = M$_2$/M$_1$) as a function of primary star mass and as a function of
companion separation.   Are (and if so, how are) companion mass ratio
statistics related to the mass distribution of single stars?

For studies of stellar populations in young clusters, the need for spatial 
resolution is dictated by crowding which limits one's ability first to
identify cluster members and then to obtain photometry at the requisite level of
accuracy ($<$2-5\%).  These issues are particularly prescient in the sub-stellar mass regime
where the objects of interest are 10-20 magnitudes fainter than the massive stars which
cohabitate the cluster.  Why might one choose to work in such difficult 
(due to point source crowding and high nebular backgrounds) regions?  The
reason is, simply, that the IMF statistics per unit angular area are far superior
in regions like the Orion Nebula Cluster, NGC 3603, or R 136 than in less observationally
complicated regions like Taurus, Ophiuchus, Lupus, or Chamaeleon.

For studies of fundamental binary properties, in particular determination of fundamental
stellar masses, the need for spatial resolution is dictated by the semi-major axis
regime one wants to probe.  For G stars the peak in the {\it stellar} companion
separation distribution occurs
around 30 AU, within the bounds of our own planetary system, which is $\sim$0.2 arc sec
at the distance of the nearest star-forming regions.  The current generation of 
large apertures (8-10 m) when used with adaptive optics enabling 
diffraction limited imaging can resolve equal-mass systems with separation 
$>$7-9 AU. Of particular interest
is the spatial resolution of known spectroscopic binaries for which combined
radial velocity and astrometric orbital monitoring leads to fundamentally determined
stellar masses.  In addition to the interesting science of understanding the
orbital (a, e) and mass ratio parameter space occupied at young ages, and hence the
observations binary formation theories must reproduce, these systems
are critical for the observational calibration of pre-main sequence evolutionary
tracks which are often used ``blindly" in IMF studies of star clusters.

\underline{Circumstellar Disks}
The topic of disks and planet formation within disks has received a lot of attention
at this meeting.  The need for high angular resolution in this area
is to obtain spatially resolved direct images of disks at
various evolutionary stages.  Measurement of disk size versus wavelength in combination
with flux versus wavelength (the spectral energy distribution) is needed in order to
correctly model these systems using absorption/re-emission and scattering codes.  Some
of the larger and brighter
disk/envelope systems have been imaged in scattered light with various
HST instruments (ACS, WFPC2, NICMOS) as well as with Keck and VLT.  
In particular we 
would like to image the innermost regions of accretion disks which serve 
both as the conduit linking material inflowing through the disk to the star, 
and as the launching point for the ubiquitous jets and outflows seen 
most prominently at the earliest evolutionary stages. Thus far
several interferometric studies have made inroads into this heretofore elusive
disk region, potentially resolving inner holes or rims/walls in disks.
Imaging of the fine structure in disks, for example gaps, warps, and other
features predicted to be a consequence of planet formation, is also high on the agenda. 
Finally, information on disk composition gradients
requires spatially resolved dust and gas spectroscopy.  We should be mindful of the
time domain when thinking about high spatial resolution imaging; inner disk and
outflow regions are known to be time-variable as might be expected given the short
dynamical time scales involved on small geometric scales.  Extended
sources in which surface brightness variability has been observed
include several disks and many jets.  Temporally resolved {\it and} spatially
resolved data on these regions are the likely path to understanding their physics.

\underline{Planet detection}
The detection of exo-solar planets is one of the great triumphs of the latter part
of the 20'th century.  After nearly a decade of discovery, our questions in this
area have become more refined and consequently more challenging in terms of the
observations needed to address them.  Future planet detection, characterization, 
and statistical analyses can be accomplished through several means including
direct (coronographic) imaging, continuation of the successful radial velocity surveys,
astrometric techniques, and transit photometry.  I mention these techniques here since
while several of them are already considered high angular resolution astronomy 
(coronographic imaging and high precision astrometry), others are in fact proxies
for high angular resolution (high spectral dispersion a.k.a. high velocity
resolution, and high photometric precision both of which when done as 
time series sample short dynamical time scales) since they are capable of 
sampling planetary system size scales on planetary system time scales.  These 
techniques offer the potential to detect planets directly (high contrast imaging, 
in/out of transit spectroscopy) or indirectly (velocity wobble, astrometric wobble, 
inference of planet-induced gaps in disks).

\section{Paths to the Future at High Angular Resolution}

As we discuss the present observational capabilities and compare them with those
available in the future, it is useful to recall that 
the angular and corresponding linear scales at the distance
of the nearest star-forming regions (140 pc) are:
1" = 140 AU, 0.1" = 14 AU, 10 mas = 1.4 AU, and 1 mas = 0.14 AU.  All of these 
scales are of interest in studies of circumstellar disk physics and planet 
formation, and are useful in studies of the initial mass function 
and companion star statistics.  
The diffraction limit at 2 $\mu$m of the current generation of premier 
optical/infrared ground-based facilities is $\sim$50 mas which will improve 
to $\sim$16 mas when the next generation of 30 m
technology is realized.  Interferometric capabilities are currently better than
5 mas fringe spacing with the smaller facilities and 3-10 mas with 
the higher sensitivity Keck and VLT Interferometers;
visibility measurements enable inference of source sizes $<$1 mas.    
Of note is the difference in field of view between the single 
apertures (several to tens of arc minutes) and the interferometers 
($<$0.1 - 1 arc seconds).  For high dispersion spectroscopy as a proxy for 
high angular resolution, under the keplerian assumption disk kinematics can be
probed (with the right emission or absorption line) at a radius inversely 
proportional to velocity resolution squared.  Even higher resolution,
R $\ga$ 10$^5$, is needed to discern finer structure in disks, such as gaps 
due to recent planet formation.  The high precision photometric monitoring 
proxy for high angular resolution in principle offers infinitely small 
angular resolution.

So what does the future hold for the optical/infrared at high angular
resolution?  I have chosen to break up the discussion into 
several time-delimited segments, depending on the wait for the capability 
at hand to emerge for main stream science usage.  
My apologies to anyone whose mission or
mission concept I do not mention.  The number of relevant facilities is 
impressive but also daunting, and I can not possibly describe them all
in the time/space allotted.

\subsection{Near-term Future Capabilities}

My near-term discussion encompasses everything from those facilities
just coming on line to those happening within the decade, 5-7 years out.

{\it Ground-Based Interferometers: KI, VLTI, and LBTI.
}
Separate papers in these proceedings discuss the recently commissioned
Keck Interferometer (KI) and Very Large Telescope Interferometer (VLTI).  
Fringes with the two 10 m Kecks (85 m baseline) on Mauna Kea were first 
obtained at 2 $\mu$m in 2001, March but operation of the full array including 
four 1.8 m outrigger telescopes is as yet uncertain.  
The interferometer will operate in V$^2$ mode (standard fringe constrast) 
at 2 $\mu$m and in nulling mode (utilizing phase inversion in one arm to
effectively null the central star) at 10 $\mu$m.  Full-array imaging 
resolution is expected to be $\sim$5 mas with 10-30 $\mu$as astrometric 
accuracy.  The VLTI at Paranal saw first fringes with two Unit telescopes
at 10 $\mu$m in July 2002 and at 2 $\mu$m in July 2003.  It is scheduled
to operate with all four 8 m Unit telescopes and 5 auxiliary telescopes
(for a total of 28 baselines).  Imaging size is 4 mas and
astrometric accuracy 10 $\mu$as.
The Large Binocular Telescope Interferometer (LBTI) is a mid-infrared 
nulling interferometer being built for operations in 2006 on Mt Graham.
Important science goals of these interferometers include measurement of
the sizes of the warm inner ($<$0.1 AU) regions of circumstellar disks, 
study of exo-zodiacal ($\sim$1 AU) dust via 10 $\mu$m nulling, 
direct detection of brown dwarfs and 
Jupiter-mass planets on close orbits to young and old stars, 
indirect detection of Uranus mass planets, and study of pre-main sequence
binary orbits for the purpose of determining fundamental stellar masses.

{\it
Space-Based Interferometers: SIM and Gaia.
}
In addition to their ground-based pursuits, both the United States and 
the European nations are planning interferometric missions in space.  
ESA's Gaia consists of three 1.4 m telescopes, two of which are used for
astrometry and one of which is for spectrophotometric data acquisition.
Launch is scheduled for 2010 into an L2 orbit with an expected 5 year
lifetime of all-sky scanning. 
Gaia's main goals are proper motions and parallaxes for 10$^9$ stars
with 4 $\mu$as precision that will connect us to the inertial reference frame.
A planet search is a by-product of the Gaia program with main operations 
revealing any Jupiter-mass or larger planet/brown-dwarf/star companions
having orbital periods between one-third and twice the mission lifetime.
NASA's Space Interferometry Mission (SIM) consists of 3 collinear
Michelson interferometers over a 10 m baseline operating at visible 
wavelengths. Astrometric accuracy will be 1 $\mu$as  
(narrow-angle) to 4 $\mu$as  (wide-angle).
Launch will be in 2009 into an Earth-trailing orbit with
operations lasting 5-10 years.  SIM is a pointed mission, surveying
2000 nearby stars, including about 150 young stars where orbital migration
might be observed amongst the sample, for astrometric signatures of massive 
planetary companions (Jupiter's pull on the Sun is 500 $\mu$as at 10 pc)
and 250 nearby stars for terrestrial-mass planets 
(Earth's pull on the Sun is 0.3 $\mu$as at 10 pc).
While radial velocity surveys can detect massive Jupiters, astrometry is 
{\it required} in order to find Earths.
Additional general science of the space interferometers includes improved 
accuracy in stellar positions, distances, proper motions, measurement of 
star cluster dynamics, and study of Galactic disk and bulge kinematics. 

{\it
Transit Missions: MOST, COROT, Kepler, and Eddington.
}
Continuing the planet-finding theme, a number of space-based efforts 
are underway to find planets through high precision photometric monitoring.  
This search method is also being pursued from the ground by a large number
of groups.  As noted above, high precision photometric monitoring is a form
of high spatial resolution given the small size scales probed.
The photometric accuracy required to detect Jupiters is 1\%,
to detect Uranus or Neptune analogs 0.1\%, and to detect Earths 0.01\%;
these constraints limit the targets to 6-16'th magnitude stars.
The space missions all operate at optical wavelengths and include
CSA's MOST launched in 2003, France's COROT scheduled for 2005, 
NASA's Kepler in development for 2007, 
and ESA's Eddington 
scheduled for 2008.
Mission goals for these spacecraft include transit detection (hundreds of
thousands of stars searched) and asteroseismology (tens of thousands of
stars searched) with studies of stellar rotation, stellar activity,
and particular classes of binary stars important ancillary science. 
The capabilities of this suite of missions are complementary. 

{\it
``Finder" Missions: SIRTF,  SOFIA,  Astro-F/IRIS,  and WISE.
}
While our focus here has been on high angular resolution, we should not 
forget to mention the suite of new low angular resolution capabilities that
in part serve as the ``finder" telescopes that will produce the targets 
to be pursued with the next generation of high angular resolution missions.  

SIRTF (Space InfaRared Telescope Facility) was launched by NASA in 2003, 
as this manuscript was being written in fact.  SIRTF
is a 0.85 m telescope in a heliocentric Earth-trailing orbit which feeds
three instruments covering 3-180 $\mu$m in eight photometric bands and
the 5-40 $\mu$m range with R=100/600 spectroscopy.  SIRTF's mission will last 
2.5-5 years.
SOFIA (Stratospheric Observatory for Infrared Astronomy) is US-German venture
consisting of a 2.7 m telescope mounted in an airborne 747SP flying 
at 39,000-45,000 feet.  The telescope is optimized for mid- to far-infrared 
operations (diffraction-limited beyond 10-15 $\mu$m) and
the instrument suite covers the full spectral range from 0.3-1600 $\mu$m at
a variety of imaging bands and with varied spectral resolution from R=1000
to $>$10,000.  First light is expected in 2005 with a 20-25 year lifetime.
SOFIA's sensitivity is intermediate between those of IRAS and SIRTF, 
but with spatial resolution a factor of 3-5 better than SIRTF.  Goals common
to both SIRTF and SOFIA include studies of star-forming regions,
searches for young super-planets and 
brown dwarfs, detailed study of proto-planetary and debris disks, and
assessment of the dust and gas disk frequency with age.  Particularly exciting 
on these missions are the spectroscopic capabilities which will give us 
a detailed look at the mineralogy and rich chemistry in disks.

Japan's Astro-F or IRIS (InfraRed Imaging Surveyor) is a 0.7 m telescope
in a polar orbit planned for around 2006.  Over 2.5-5 years, 
with one instrument it will conduct a low-resolution, 30-50", all-sky survey 
at 50 and 200 $\mu$m while with a second instrument it will obtain
pointed observations at higher resolution, 2", through filters
ranging from 2-26 $\mu$m.   WISE (Wide-field 
Infrared Survey Explorer) or NGSS as it was formerly known, is a proposed
4-band 3.5-23 $\mu$m all-sky survey at 5-10" resolution for launch by NASA
in 2008 to a polar orbit.  On a 6-12 month mission this 0.5 m telescope will 
reach all-sky sensitivities in the mid-infrared comparable to those 
of the optical SDSS.

SIRTF and SOFIA are pointed observatories while Astro-F and WISE are
all-sky surveys.  Collectively, these relatively low spatial resolution
infrared missions will provide a wealth of targets for follow-up study
with the next generation of high angular resolution facilities where, for
example, newly found companions to Sun-like stars can be characterized and 
newly discovered disks can have their
composition and structure as a function of disk radius understood.

\subsection{Mid-term Future Capabilities}

These are the beneficiaries of the aforementioned ``finder" missions.
I have only one space-based mission and some interesting prospects for the 
ground to discuss in this category of 8-15 years hence facilities.

The next generation space telescope (NGST) has been re-named JWST and is
presently a 5.6 m segmented mirror telescope.   JWST will be passively cooled 
at its L2 orbit, feeding three instruments that will
obtain photometry and (multi-object) spectroscopy at resolution several 
hundred to several thousand.  Covering 1-28 $\mu$m the near-infrared field 
of view is 5' and the mid-infrared 1.5'.
Launch is presently scheduled for 2011 with a 5-10 year mission anticipated.
The detectors will suffer backgrounds 6 orders of magnitude lower in
the thermal infrared compared to ground-based facilities and will achieve 
nano-Jansky (1-3 $\mu$m near-infrared) to better than micro-Jansky 
(10-20 $\mu$m) sensitivities.
JWST's objectives are to detect the ``first (near-infrared) light" from
primordial stars and galaxies and to measure the creation of the first
heavy elements.  However, it will also have a strong program of general
astrophysics related to star and planet formation including measurement
of the stellar/sub-stellar mass function in extreme star-forming environments,
determination of the minimum mass that can freely fragment from a 
molecular cloud, and studies of gas and dust disks around young stars.
JWST is a NASA, ESA, and CSA partnership.

Now what about the ground?  Our premier large-aperture ground-based facilities
(Keck I and II at 10m, Subaru at 8m, Very Large Telescopes 1-4 and
Gemini North and South all at 8m, and the forthcoming South African Large
Telescope at 11 m) are envisioned to lead to a next generation of
telescopes that may be 15-20 m (e.g. the single aperture LAT and VLOT or the
multiple aperture 20-20 concepts) to 30 m (CELT or GSMT) or even larger 
50-100 m (Euro50 or OWL) in size.   Of great amusement is the comparison of
one of these behemoths to a modern day sports stadium where the aperture and
supporting structure takes up the majority of a rugby, soccer, or football field
and the telescope enclosure is nearly the size of the entire stadium.  We have 
quite a bit of technology development and ingenious engineering on the
large telescope design and development agenda.  

Speaking
more seriously now about the comparative advantages of the ground and of space, it is
important to note that there are complementary roles to be played by a 20-30 m 
ground-based facility and a 5-6 m space-based facility.  For imaging, 
while the near-infrared (1-2.5 $\mu$m) sensitivities of, as examples,
a 30 m ground-based telescope and a 6 m space-based telescope 
are similar in the background-limited case, 
with diffraction-limited performance the 30 m has better spatial resolution by the
ratio of the aperture sizes, or approximately a factor of 5.   Since we will always build
bigger light buckets on the ground than we send into space, this advantage should be
maintained.    In the mid-infrared ($>$3 $\mu$m) ground-based sensitivity is not
competitive with that available from space.  However, as mentioned above, the ground-based
spatial resolution will always be better in the diffraction-limited case. 
Further, other capabilities such as high dispersion mid-infrared spectroscopy 
may become available only from the ground. For near-infrared spectroscopy, 
at moderate spectral resolution R$\approx$4000, the ``air glow" due mainly 
to OH emission lines is suppressed in the red optical and near-infrared
which means that traditional backgrounds are reduced (by a factor of $\sim$30)
making the ground competitive with space.  Further, truly high dispersion spectrographs
or other specialized capabilities such as integral field units
are unlikely to be built as space instruments, defining yet other niches for the ground.
These trades between wavelength and spectral resolution have been quantified 
by the CELT and GSMT projects.  Great complementarity is expected between the
next generation of 30 m ground and 5-6 m space facilities from 1-25 $\mu$m.

\subsection{Far-term Future Capabilities}

It is never too early to plan for the future according to some financial
advisors, and the same may be said for big-ticket space facilities paid for
with public money.  In the era of longterm strategic planning that is
presently fashionable in science, there is always
an identified horizon filled with the necessary as well as the exotica.  
Mission concepts which have emerged thus far in the $>$15-20 year time frame
are a combination of both.

NASA's Terrestrial Planet Finder (TPF) and ESA's Darwin mission both have
as their primary goal the discovery and characterization of Earth-like,
potentially habitable planets.  TPF is being studied as either an infrared
(7-20 $\mu$m) separated spacecraft interferometer achieving nulling to one part
in 10$^6$ and detecting planets via their thermal emission, or an optical 
wavelength coronograph expecting starlight suppression to one part in 10$^9$  
and detecting planets via reflected light from the central star.
TPF is aiming for 2015 launch into either L2 or an Earth-trailing orbit.
Darwin is cast as a six-element infrared interferometer plus central
beam combiner planned for 2015 or later and also located at L2.
The potential for collaboration between these missions is to be encouraged, 
given their great expense and vast scientific overlap.
Achieving the first of their goals means that a statistically significant 
number of stars must be surveyed and the observations must be capable of 
discovering Earths located in the ``habitable zone" of liquid water existence.  
The second goal, characterization of Earth-like planets requires a search
for atmospheres, habitability, and perhaps even biosignatures through 
indicators such as CO$_2$, H$_2$O, N$_2$O, O$_2$, O$_3$, and CH$_4$; these
molecules in particular were all seen by the Mars Express Mission as it 
looked back and took a spectrum of Earth just before this meeting (July 2003). 
For more general astrophysics purposes, it should be appreciated 
that these missions will have spatial resolution 10-100 times that of JWST
and are thus capable of far more star and planet formation science than just
their articulated primary goals.

At short wavelengths, the Space UltraViolet Optical (SUVO) telescope
is a conceived successor to Hubble (HST) operating in the ultraviolet 
and visible range.  While focusing on the evolution of the first galaxies,
quasars, and stars, this observatory would also provide substantial
high spatial resolution ($<$30 mas) capability for star and planet formation 
studies over the needed wide fields ($>$10').  
A 4-8m (deployable optics) aperture is envisioned with launch hoped for
by the end of next decade.

At long wavelengths, the Single Aperture Far-InfraRed (SAFIR) observatory
and the Submillimeter Probe of Early Cosmological Structure (SPECS) 
offer our first chance for truly high spatial resolution in the far infrared.  
SAFIR is a follow-on to NGST, SIRTF/Herschel, and SOFIA, 
an 8-10 m telescope operating from $\sim$20-500$+\mu$m (diffraction-limited
beyond 40 $\mu$m) with both imaging and R$\approx$1000 spectroscopy.  
Both filled and sparse-aperture mirrors are being explored.
The facility is expected to be limited only by astrophysical backgrounds
(zodiacal light, galactic cirrus, and the cosmic microwave background) 
and thus will achieve heretofore unrealized sensitivity in the mid- and 
far-infrared.  Launch is contemplated for the 2015-2020 time frame to L2.
SPECS intends to operate several 
1 km baselines and achieve HST-like resolution (50 mas) in the far-infrared,
a truly remarkable capability, if realized.
Science goals of SAFIR and SPECS include study of the interstellar medium 
in star-forming regions, protostellar objects, and the more evolved
proto-planetary and debris disks.

\subsection{Dreamland} 

There are a few concepts even further afield than the somewhat mature but still
developing and evolving mission concepts
outlined above.  Planet Imager and Life Finder are two in the NASA 
long term plan, with launch dates approaching 2050.  Planet Imager hopes to
obtain spatially resolved images of planets orbiting other stars while
Life Finder is a spectroscopy mission geared towards finding signatures
of life (biomarkers).

\section{Concluding Remarks}

As reviewed in this talk, there are a variety of high angular resolution 
techniques or proxies thereof under development, or just coming 
online now, that are of interest for star and planet formation studies.  
Direct imaging with large apertures, interferometry, high dispersion 
spectroscopy, and transit photometry are among the maturing technologies.
A series of ground-based telescope concepts and space mission concepts 
are waiting in line over the next several decades for their turn
to implement these technologies and techniques.
How do we decide as a community what capabilities to pursue?  In order to 
secure either public or private funding at the scale needed to carry out 
our current plans and dreams
there must exist an alignment of the following four concepts:
\begin{itemize}
\item important science questions
\item technological readiness
\item cost
\item correspondence with strategic goals
\end{itemize}
We astronomers are fortunate to live in a time when this is not only possible,
but desired and expected by the science-consuming public at large.  
Complementary science will be done from the ground and from space with the
next generation facilities.  Another general conclusion from the present survey 
of the future is that L2 has the potential to become quite crowded 
by about 2020!

\acknowledgements
I am grateful to the following for providing some of the slide material used
in this talk: Mike Devirian, Mark McCaughrean, Richard Ellis, Alycia Weinberger
and of course the WWW at large.   I also wish to commend Hans Zinnecker for his
statement during the conference summary talk about a talk not being a paper; 
in this instance even a conference talk proceedings 
is not a comprehensive paper on the topic at hand!  Finally, my gratitude to the
conference organizers for the invitation to experience Oz.
 
%

\end{document}